\documentclass[prb,twocolumn,showpacs]{revtex4}
\usepackage{graphicx}
\usepackage{amsmath}
\usepackage{amssymb}
\usepackage{dcolumn}
\usepackage{float}
\usepackage{bm}

\renewcommand{\Re}{\mathrm{Re}\,}
\renewcommand{\Im}{\mathrm{Im}\,}

\newcommand{\p}{\partial}

\renewcommand{\Re}{\mathrm{Re}\,}
\renewcommand{\Im}{\mathrm{Im}\,}

\DeclareMathAlphabet{\bi}{OML}{cmm}{b}{it}

\def\be{\begin{equation}}
\def\ee{\end{equation}}
\def\bearr{\begin{eqnarray}}
\def\eearr{\end{eqnarray}}
\def\la{\langle}
\def\ra{\rangle}

\def\bs{\boldsymbol}

\begin{document}
\title{Wave packet dynamics and zitterbewegung of heavy holes in
a quantizing magnetic field}
\bigskip

\author{Tutul Biswas and Tarun Kanti Ghosh}
\normalsize
\affiliation
{Department of Physics, Indian Institute of Technology-Kanpur,
Kanpur-208 016, India}
\date{\today}
 
\begin{abstract}
In this work we study wave packet dynamics and  
$zitterbewegung$, an oscillatory quantum motion, of heavy 
holes in III-V semiconductor quantum wells in presence of a 
quantizing magnetic field. It is revealed
that a Gaussian wave-packet describing a heavy hole  
diffuses asymmetrically along the circular orbit while performing cyclotron 
motion. The wave packet splits into two peaks with unequal amplitudes after 
a certain time depending on spin-orbit coupling constant. 
This unequal splitting of the wave packet is attributed 
to the cubic Rashba interaction for heavy holes. 
The difference in the peak amplitudes disappears with time.
At a certain time the two peaks diffuse almost along the entire cyclotron 
orbit. Then tail and head of the 
diffused wave packet interfere and as a result a completely randomized pattern of 
the wave packet is observed. 
The diffusion rate of the wave packet increases with increase of the
spin-orbit interaction strength. Also strong spin-orbit coupling expedite 
the splitting and the randomization
of the wave packet. We also study the $zitterbewegung$ in various 
physical observables such as position, charge current and spin angular momentum
of the heavy hole. The $zitterbewegung$ oscillations are very much sensitive to the 
initial wave vector of the Gaussian wave packet
and the strength of the Rashba spin-orbit coupling.

\end{abstract}



\maketitle

\section{Introduction}

Spin dependent transport phenomena in low-dimensional semiconductor 
structures have been of a lot of interest to the scientific community 
in recent years due to the potential applications in the highly 
emerging field of spintronics.\cite{winkler,fabian,zutic,cahay}
Intense research in this field was initiated after the proposal of spin 
field effect transistor by Datta and Das \cite{datta}. The principal aim 
of this field is to produce pure spin current and its manipulation 
on semiconductor nanostructure devices. One important tool for generating 
pure spin current is the well known spin Hall effect (SHE). 
\cite{hirsch,zhang,murakami,sinova,hankie,kato,sheng,wund}
In SHE spin-orbit interaction (SOI) leads to a generation of spin current from 
an external electric voltage. For the happening of SHE, the charge carriers 
are either electrons in the conduction band or holes in the valence band
in a III-V semiconductor such as GaAs. 
Many $p$-doped semiconductors such as GaAs, InSb, Si etc show four-fold 
degeneracy in their valence band around the $\Gamma$ point. This kind 
of systems is described by $4\times4$ Luttinger\cite{Lutt} 
Hamiltonian. 
The Luttinger Hamiltonian leads to two two-fold degenerate 
energy branches having different effective masses. These two 
dispersion branches are known as the heavy-hole (HH) and light-hole (LH) 
bands which are described by $j_z = \pm 3/2$ and $j_z = \pm 1/2$, respectively. 
For spin-orbit coupled two-dimensional hole system (2DHS) in a narrow 
quantum well, the $4\times4$ Luttinger Hamiltonian can be projected onto 
$|3/2,\pm 3/2\ra$ HH states giving rise to an effective $2\times2$
Rashba Hamiltonian\cite{sczhang,rwink,loss} provided that the 
hole density is sufficiently low. In 2DHS the Rashba SOI is tri-linear in momentum.

In recent years an oscillatory quantum motion, called $zitterbewegung$ (ZB), 
has become a central point of research interest in various low-dimensional 
semiconducting systems. 
The length and time scales corresponding to this trembling motion 
in vacuum are of the order of the Compton wavelength 
$\lambda_c = \hbar/m_0c\sim10^{-4}$ nm 
and $\hbar/(2m_0c^2) \sim 10^{-22}$ s, respectively. Due to this ultra small 
length and time scale, it has not been possible to make an experimental verification 
of ZB phenomenon in vacuum so far. However an intense interest was initiated 
around 2005 by Zawadzki\cite{zawad1} who has drawn an analogy between the Dirac 
equation of a free Dirac electron and the ${\bf k}\cdot{\bf p}$ 
theory of the band structure of a narrow gap semiconductor. He was 
able to find out a new length scale $\lambda_Z \approx 7$ nm which is much
larger than $\lambda_c$. Later, the problem of ZB of electrons and holes 
in III-V semiconductor quantum wells in the presence of spin-orbit interaction 
was studied by Schliemann\cite{schliem1,schliem2} et. al. These
pioneering works initiated tremendous motivation for theoretical studies in search
of ZB in various condensed matter systems 
such as crystalline solid,\cite{tmrush} 
graphene,\cite{zawad2,rusin,demi1,yang,zawadzr,jung1,jung2,rakhi,costa}
carbon nanotubes,\cite{zawad2,zawad3,zawad5} Luttinger liquid,\cite{demi2}
superconductor,\cite{cserti1} ultra-cold atom,\cite{vaishnav,song} 
topological insulators \cite{chang1,chang2} etc. It was also reported that 
the origin of the minimal conductivity \cite{kats} in graphene can be explained 
in the light of the peculiar phenomenon ZB.
Recently, a general theory for ZB 
has been developed by David and Cserti.\cite{cserti2} 
Winkler\cite{winkler2} et. al. considered a number of effective
Hamiltonians representing different systems and studied various consequences
of ZB oscillations. The effect of an in-plane magnetic field on the ZB 
oscillations in a Rashba-Dresselhaus system has been studied.\cite{ghosh}
Very recently, a complex quantum motion known as ``super zitterbewegung'' in graphene
has been studied\cite{rushin5} theoretically using time-dependent two-band Hamiltonian in the
framework of rotating wave approximation.

Although the ZB phenomenon in vacuum is not yet observed experimentally, 
however an optical and acoustic analog of relativistic ZB have been observed 
experimentally in optical super-lattice\cite{felix} and in a 
two-dimensional sonic crystal\cite{liu}, respectively. 
Also simulation\cite{gerrit} of Dirac particles has been performed
recently using trapped ions and laser excitations. 
Most recently, the ZB in $^{\bf87}$Rb Bose-Einstein condensate is observed
experimentally \cite{spielman,Engels} using direct imaging 
technique.

In this work we consider the long standing problem of ZB of 
heavy holes with cubic Rashba interaction in a III-V semiconductor 
quantum well subjected to a perpendicular magnetic field.
We study time-evolution of a heavy hole represented by the Gaussian 
wave packet. 
We visualize and discuss how the hole wave packet evolves with 
time around the cyclotron orbit. It is shown that as time goes on, the 
initial wave packet starts to diffuse asymmetrically along the cyclotron orbit 
while
executing cyclotron motion. At a later time, depending on the spin-orbit 
strength, the wave packet splits into two peaks with unequal amplitudes. 
After many more cycles, the wave packet diffuses entirely along the
cyclotron orbit. Then interference occurs between tail and head of the 
diffused wave packet, which randomizes the wave packet along the entire 
cyclotron orbit. 
The ZB of heavy holes in various physical quantities, such as position, charge current and 
spin angular momentum, are studied analytically as well as numerically. 
It is revealed that the ZB in these observables are very much dependent on 
the initial wave vector of the Gaussian wave packet and the Rashba 
spin-orbit interaction.

This paper is organized in the following way. 
In section II we present all relevant theoretical details 
such as the Hamiltonian, Landau levels, its corresponding eigenfunction
and time-evolution of initial 
hole wave packet using Green's function approach as well as 
average values of various physical observables. Numerical
results and discussion are given in 
section III. We summarize our findings in section IV.
We present detailed calculations in Appendix A.

\section{Theoretical Calculations}

\subsection{Hamiltonian}

The dynamics of holes in the $\Gamma_8$ valence band of III-V semiconductors with 
zinc-blende structure like GaAs are well described within the framework of the Luttinger model (LM).
In the spherical approximation\cite{spheric} of LM, the topmost valence band is four fold degenerate
at the $\Gamma$ point with HH and LH bands corresponding to total spin $J=3/2$.
Strong quantum confinement in the semiconductor heterostructure along the growth
direction removes the $\Gamma$ point degeneracy between  
the HH and LH bands. At very low temperature and low density only
the HH subbands are assumed to be occupied.
Now, it is possible to obtain a $k$-cubic Rashba Hamiltonian 
from the Luttinger Hamiltonian by projecting the latter onto the HH states.

It is well known\cite{rwink,sczhang} that the Rashba spin-splitting for HH and 
LH bands are proportional to $k^3$ and $k$, respectively. The linear spin-splitting 
for LHs is similar to that for electrons.\cite{rashba} Our present study mainly deals 
with only HHs and hence linear spin-splitting for electrons and LHs is 
completely ignored.

In presence of a perpendicular magnetic field the 
single particle Hamiltonian \cite{tianx,zarea} of a 2DHS with Rashba SOI (RSOI) 
can be written as 
\begin{eqnarray}\label{Ham1}
H=\frac{{\bf \Pi}^2}{2m^\ast}+\frac{i \alpha}{2\hbar^3}
\Big(\Pi_{-}^3\sigma_{+}-\Pi_{+}^3\sigma_{-}\Big)
- g_s \mu_B {\bf J} \cdot {\bf B},
\end{eqnarray}
where 
${\bf \Pi}={\bf p}-e{\bf A}$ with ${\bf p}$ is the
canonical momentum operator, ${\bf A}$ is the vector potential corresponding
to the external magnetic field ${\bf B} $, ${\bf J}=(3/2){\bs \sigma}$,
$m^\ast$ is the effective mass of the
heavy hole, $\alpha$ is the Rashba coupling coefficient and $\sigma_i$ 's are the
Pauli matrices.
Also,
$\Pi_{\pm}=\Pi_x\pm i\Pi_y$, $\sigma_\pm=\sigma_x\pm i\sigma_y$ and 
$g_s$ is the effective Lande g-factor. 
One important point to be mentioned 
here that the Pauli matrices represent an effective pseudo-spin with 
spin projection $\pm3/2$ along the growth direction of the quantum well.

For convenience, we assume $ {\bf B} = - B \hat{z} $ and 
the corresponding vector potential in the Landau gauge as 
${\bf A}=(By,0,0)$.
The Hamiltonian $H$ in Eq. (\ref{Ham1}) commutes with $p_x$. 
Hence the wave vector $k_x$ is a good quantum number.
In matrix form the Hamiltonian takes the following form: 
\begin{eqnarray}\label{Ham3}
H=\hbar\omega_c
\begin{pmatrix}
h_0 + \chi & -i \tilde{\alpha}h_{-}^3 \\
i \tilde{\alpha} h_{+}^3 &  h_0 - \chi\\
\end{pmatrix},
\end{eqnarray}
where $h_0=(-\p^2/\p\eta^2+\eta^2)/2 $ is the oscillator Hamiltonian 
and $ h_{\pm}= (\eta\mp\p/\p\eta)/\sqrt{2}$ is
the creation (annihilation) operator. Here, $ \eta=(y-y_c)/l_0 $ with
$y_c = l_0^2k_x $ and
$l_0=\sqrt{\hbar/(eB)}$ is the magnetic length.
Also, $\omega_c=eB/m^\ast$ is the cyclotron frequency, the 
dimensionless Rashba parameter 
$\tilde{\alpha}=2\sqrt{2}\alpha/(l_0^3\hbar\omega_c)$ and
$ \chi = 3g_sm^*/(4m_e)$.
The operations of $h_0$ and $h_\pm $ on the oscillator wave functions
$ \phi_n(\eta) = \sqrt{1/(2^nn!\sqrt{\pi} l_0)} e^{-\eta^2/2}
H_n(\eta)$  
are $h_0\phi_n(\eta)=(n+1/2)\phi_n(\eta)$, 
$h_{+}\phi_n(\eta)=\sqrt{n+1}\phi_{n+1}(\eta)$ 
and $h_{-}\phi_n(\eta)=\sqrt{n}\phi_{n-1}(\eta)$.

For $n\geq3$ the energy eigenvalues\cite{tianx} are
given by 
\begin{eqnarray}\label{eigen}
\epsilon_n^\lambda= \hbar\omega_c \Big[(n-1)+\lambda
\sqrt{E_0^2
+\tilde{\alpha}^2A_n} \Big],
\end{eqnarray}
where $\lambda=\pm$ represents two spin-split energy branches, 
$E_0=\chi - 3/2$ and $A_n=n(n-1)(n-2)$.
The eigenfunctions corresponding to the eigenvalues given 
in Eq. (\ref{eigen}) are
\begin{eqnarray}\label{wavefnn}
\psi_{\xi}^\lambda({x,y})=\frac{e^{ik_xx}}{\sqrt{2\pi}}\left(
\begin{array}{c}
i\cos\theta_n^\lambda\phi_{n-3}(y-y_c)\\
\sin\theta_n^\lambda\phi_n(y-y_c)\\
\end{array}\right).
\end{eqnarray}
Here, $\xi: \{n,k_x\} $ is a set of two quantum numbers, 
$\tan\theta_n^\lambda=\tilde{\alpha}\sqrt{A_n}/B_n^\lambda$
with $B_n^\lambda=n+1/2-\chi-\epsilon_n^\lambda/(\hbar\omega_c)$.

For $n<3$, the energy levels are 
$ \epsilon_n^{+}=\hbar\omega_c (n + 1/2 - \chi) $
and the corresponding eigenfunctions are
\begin{eqnarray}\label{wavefn0}
 \psi_{\xi}^+({x,y})=\frac{e^{ik_xx}}{\sqrt{2\pi}}\left(
\begin{array}{c}
0\\
\phi_n(y-y_c)\\
\end{array}\right).  
\end{eqnarray}

\subsection{Propagator construction and time-evolution of 
initial wave packet}

In order to describe the time-evolution of an initial Gaussian 
wave packet it is necessary to construct an appropriate propagator. 
For this purpose we follow the well known Green's function technique 
as described in Ref.~[\onlinecite{demi}].
In matrix form, the propagator or the Green's function can be written as 
\begin{eqnarray}\label{GreenM}
G({\bf r}, {\bf r^\prime},t)=
\begin{pmatrix}
G_{11}({\bf r}, {\bf r^\prime},t) & G_{12}({\bf r}, {\bf r^\prime},t) \\
G_{21}({\bf r}, {\bf r^\prime},t)&  G_{22}({\bf r}, {\bf r^\prime},t)\\
\end{pmatrix}.
\end{eqnarray}
The matrix elements of Eq. (\ref{GreenM}) are defined as 
\begin{eqnarray}
G_{ij}({\bf r}, {\bf r^\prime},t)=\sum_{\lambda=\pm}
\int dk_x\sum_{n=0}^{\infty}\psi_{\xi,i}^\lambda({\bf r},t)
\psi_{\xi,j}^{\lambda^\ast}({\bf r}^\prime,0).
\end{eqnarray}
Here, the indices $i,j=1,2$ represent the upper/lower components of the 
wave functions.
The hole wave function at a later time $t$ is given by 
$\psi_\xi^\lambda({\bf r},t)=\psi_\xi^\lambda({\bf r},0) 
e^{-i\epsilon_n^\lambda t/\hbar}$, 
where $\psi_\xi^\lambda({\bf r},0)$ 's are given by 
Eqs. (\ref{wavefnn}) and (\ref{wavefn0}).

In presence of the perpendicular magnetic field,
we represent an initial state of the hole by 
a Gaussian wave packet with the initial spin polarization along
the $z$ axis as given by 
\begin{eqnarray}\label{initial}
\Psi({\bf r},0)=\frac{1}{\sqrt{\pi}l_0}
\exp\Big(-\frac{r^2}{2l_0^2}+i\frac{p_{0x}x}{\hbar}\Big)\left(
\begin{array}{c}
1\\
0\\
\end{array}\right),
\end{eqnarray}
where $p_{0x} = \hbar k_{0x}$ is the initial momentum of the 
wave packet along $x$ direction.
Note that the initial state coincides with the coherent state of a charge
particle in a magnetic field. We have taken such wave function because
the dynamics of coherent states in a magnetic field resembles the dynamics of a 
classical particle.

The choice of the two component spinor considered in Eq. (\ref{initial})
is completely arbitrary. It can have 
both nonzero components as discussed in Refs.~[\onlinecite{rusin,demi1,zawad5}].

In general the time-evolution of the initial wave packet can be
obtained with the help of the Green's function as given by
\begin{eqnarray}\label{time-evo}
\Psi({\bf r},t)=\int d{\bf r}^\prime G({\bf r},{\bf r}^\prime,t)
\Psi({\bf r}^\prime,0).
\end{eqnarray}

Using Eq. (\ref{GreenM}), (\ref{initial}) and (\ref{time-evo}) one can obtain 
\begin{eqnarray}
\left(
\begin{array}{c}
\Psi_1({\bf r},t)\\
\Psi_2({\bf r},t)\\
\end{array}\right)
& = & \frac{1}{\sqrt{\pi}l_0}\int d{\bf r}^\prime 
\exp\Big(-\frac{{r^\prime}^2}{2l_0^2}
+i\frac{p_{0x}x^\prime}{\hbar}\Big) \nonumber \\ 
& \times &
\left(
\begin{array}{c}
G_{11}({\bf r}, {\bf r}^{\prime}, t)\\
G_{21}({\bf r}, {\bf r}^{\prime}, t)\\
\end{array}\right).
\end{eqnarray}

It is straightforward to calculate the required matrix elements 
of the Green's function and these are given by 
\begin{eqnarray}
G_{11}({\bf r}, {\bf r}^\prime,t)&=&\frac{1}{2\pi}\int_{-\infty}^{\infty}
dk_xe^{ik_x(x-x^\prime)}\sum_{n=0}^\infty\Gamma_{n+3}(t)\nonumber\\
&\times&\phi_n(y-y_c)\phi_n(y^\prime-y_c)
\end{eqnarray}
and 
\begin{eqnarray}
G_{21}({\bf r}, {\bf r}^\prime,t)&=&\frac{1}{2\pi}\int_{-\infty}^{\infty}
dk_xe^{ik_x(x-x^\prime)}\sum_{n=0}^\infty\Delta_{n+3}(t)\nonumber\\
&\times&\phi_{n+3}(y-y_c)\phi_n(y^\prime-y_c),
\end{eqnarray}
where $\Gamma_n(t) = e^{-i(n-1)\omega_ct}
[\cos(\delta_n t)-i\xi_n\sin(\delta_n t)]$
and $\Delta_n(t) = e^{-i(n-1)\omega_ct}\zeta_n\sin(\delta_nt)$ with 
$\delta_n=\omega_c\sqrt{E_0^2
+\tilde{\alpha}^2A_n}$, $\xi_n=E_0/\sqrt{E_0^2
+\tilde{\alpha}^2A_n}$ and
$\zeta_n=\tilde{\alpha}\sqrt{A_n}\xi_n/E_0$.
Note that $ \Gamma_n(t=0) = 1 $ and $ \Delta_n(t=0) =0 $. 
It is easy to verify an interesting result that 
$\vert \Gamma_n(t)\vert^2+\vert \Delta_n(t)\vert^2=1$.

The components of the wave packet at a later 
time $t$ are given by 
\begin{eqnarray}\label{wavefn1}
\Psi_1({\bf r}, t)&=&\frac{1}{\sqrt{2}\pi l_0}\sum_{n=0}^\infty
(-1)^n \frac{\Gamma_{n+3}(t)}{2^nn!}\nonumber\\
&\times&\int_{-\infty}^\infty due^{\Lambda(x,y,u)}
u^n H_n(y/l_0-u)
\end{eqnarray}
and 
\begin{eqnarray}\label{wavefn2}
\Psi_2({\bf r}, t)&=&\frac{1}{4\pi l_0}\sum_{n=0}^\infty
(-1)^n \frac{\Delta_{n+3}(t)}{2^nn!\sqrt{A_{n+3}}}\nonumber\\
&\times&\int_{-\infty}^\infty due^{\Lambda(x,y,u)}
u^n  H_{n+3}(y/l_0-u),
\end{eqnarray}
where $u=k_xl_0$ and 
$\Lambda(x,y,u)=iux/l_0-(a-u)^2/2-u^2/4-(y/l_0-u)^2/2$ with
$ a = k_{0x}l_0$.

\subsection{Zitterbewegung in position and velocity}
In this section we find the average values of the position 
and velocity operators. The time-dependent average value of 
the position operator is given by

\begin{eqnarray}\label{position}
 \left(
\begin{array}{c}
\la x\ra\\
\la y\ra\\
\end{array}\right)
=\sum_{i=1,2}\int d{\bf r}\Psi_i^\ast({\bf r},t)
\left(
\begin{array}{c}
x\\
y\\
\end{array}\right)
\Psi_{i}({\bf r},t).
\end{eqnarray}

It is straightforward to calculate the average values 
of $x$ and $y$ using Eqs. (\ref{wavefn1}) and (\ref{wavefn2}).
Detailed calculations are given in Appendix A.
The average values of $x$ and $y$ are given by
\begin{eqnarray}
\la x(t) \ra & = & L 
\sum_{n=0}^\infty\frac{i (-1)^{n+1}}{n! (12)^{n}}
\Big\{\Im(\Gamma_{n + 4}^\ast\Gamma_{n+3}) + \sqrt{\frac{n+4}{n+1}} \nonumber \\
& \times & \Im(\Delta_{n+4}^\ast\Delta_{n+3})\Big\}
H_{2n+1}(i\sqrt{2/3}a)
\end{eqnarray}
and 
\begin{eqnarray}\label{PosY}
\la y(t) \ra & = & L 
\sum_{n=0}^\infty\frac{i(-1)^{n}}{n! (12)^n}
\Big\{\Re(\Gamma_{n+4}^\ast\Gamma_{n+3}) + \sqrt{\frac{n+4}{n+1}} \nonumber\\
& \times &\Re(\Delta_{n+4}^\ast\Delta_{n+3}) - 1 \Big\}
H_{2n+1}(i\sqrt{2/3}a),
\end{eqnarray}
where $ L = \frac{l_0}{3} \exp(-a^2/3)$.

The velocity operator is obtained from the commutation relation 
${\bf v}= [{\bf r}, H]/(i \hbar)$.
The components of the velocity operator are given by
\begin{eqnarray}
v_x=\frac{\Pi_x}{m^\ast}\sigma_0+\frac{3i\alpha}{2\hbar^3}
(\sigma_{+}\Pi_{-}^2-\sigma_{-}\Pi_{+}^2)
\end{eqnarray}
and
\begin{eqnarray}
v_y=\frac{\Pi_y}{m^\ast}\sigma_0+\frac{3\alpha}{2\hbar^3}
(\sigma_{+}\Pi_{-}^2+\sigma_{-}\Pi_{+}^2).
\end{eqnarray}
The average value of the velocity operator is given by 
\begin{eqnarray}\label{vel_exp}
\la v_k(t)\ra=\int dxdy\left(
\begin{array}{c}
\Psi_1^\ast \ \Psi_2^\ast
\end{array}\right)
{v_k}
\left(
\begin{array}{c}
\Psi_1 \\ \Psi_2
\end{array}\right),
\end{eqnarray}
where the index $k$ represents the  $x$ and $y$ components
of the velocity.
Using Eqs. (\ref{wavefn1}), (\ref{wavefn2}) and (\ref{vel_exp}),
after a lengthy but straightforward calculation, we finally obtain 
average values of the components of the velocity operator as 
\begin{eqnarray}\label{vel_x}
\la v_x(t)\ra & = & L \omega_c
\sum_{n=0}^\infty\frac{i(-1)^{n+1}}{n!(12)^n} 
\Big\{\Re(\Gamma_{n+4}^\ast\Gamma_{n+3})  +  \sqrt{\frac{n+4}{n+1}} \nonumber\\
& \times & \Re(\Delta_{n+4}^\ast\Delta_{n+3})
+3\tilde{\alpha}\sqrt{\frac{(n+2)(n+3)}{n+1}}\nonumber\\
&\times&\Im(\Gamma_{n+4}^\ast\Delta_{n+3})\Big\} 
H_{2n+1}(i\sqrt{2/3}a)
\end{eqnarray}

and
\begin{eqnarray}\label{vel_y}
\la v_y(t)\ra & = & L \omega_c
\sum_{n=0}^\infty\frac{i(-1)^{n+1}}{n!(12)^n} 
\Big\{\Im(\Gamma_{n+4}^\ast\Gamma_{n+3}) + \sqrt{\frac{n+4}{n+1}} \nonumber\\
& \times & \Im(\Delta_{n+4}^\ast\Delta_{n+3})
-3\tilde{\alpha}\sqrt{\frac{(n+2)(n+3)}{n+1}}\nonumber\\
&\times&\Re(\Gamma_{n+4}^\ast\Delta_{n+3})\Big\}
H_{2n+1}(i\sqrt{2/3}a).
\end{eqnarray}

\subsection{Zitterbewegung in spin angular momentum}
Now we turn to concentrate on the ZB in the spin angular momentum
of a heavy hole in the presence of a perpendicular magnetic field. 
In this sub-section we shall calculate the time-dependent average 
values of the components of the effective spin operator 
\cite{naka,raichev} $ {\bf S}=\frac{3\hbar}{2}{\bs \sigma}$.
The spatio-temporal profile of the hole spin density is defined as  
\begin{eqnarray}\label{spin_den}
{\bf S}({\bf r},t)=
 \left(
\begin{array}{c}
\Psi_1^\ast \ \Psi_2^\ast
\end{array}\right)
{\bf S}
\left(
\begin{array}{c}
\Psi_1 \\ \Psi_2
\end{array}\right).
\end{eqnarray}

The time-dependent average value of the spin operator 
is given by
\begin{eqnarray}
\la {\bf S}(t)\ra=\int {\bf S}({\bf r},t) dxdy.
\end{eqnarray}
The components of the average values of the 
spin operators are 
\begin{eqnarray}\label{spinx}
\la S_x(t) \ra & = & \frac{\hbar}{\sqrt{2}}\exp(-a^2/3)
\sum_{n=0}^\infty \frac{i(-1)^{n+1}}{n!(12)^{n+1}} \nonumber\\
& \times &  \frac{\Re(\Gamma_{n+6}^\ast\Delta_{n+3})}{\sqrt{A_{n+3}}} 
H_{2n+3}(i\sqrt{2/3}a),
\end{eqnarray}

\begin{eqnarray}\label{spiny}
\la S_y(t) \ra & = & \frac{\hbar}{\sqrt{2}}\exp(-a^2/3)
\sum_{n=0}^\infty\frac{i(-1)^{n+1}}{n!(12)^{n+1}}\nonumber\\
& \times & \frac{\Im(\Gamma_{n+6}^\ast\Delta_{n+3})}{\sqrt{A_{n+3}}} 
H_{2n+3}(i\sqrt{2/3}a)
\end{eqnarray}
and 
\begin{eqnarray}\label{spinz}
\la S_z(t)\ra & = & \sqrt{\frac{3}{2}} \hbar \exp(-a^2/3)
\sum_{n=0}^\infty\frac{(-1)^n }{n!(12)^n}\nonumber\\
&\times& 
(\vert\Gamma_{n+3}\vert^2 -\vert\Delta_{n+3}\vert^2)
H_{2n}(i\sqrt{2/3}a).
\end{eqnarray}

\section{Numerical results and discussions}
In this section we shall visualize and discuss how 
the spatial distribution of probability density for the 
heavy holes changes with time. We shall also discuss the 
time dependencies of the expectation values of the position, 
current and spin operators and their various consequences.
For numerical calculations, we adopt the material parameters 
appropriate for GaAs quantum wells. 
We took $m^\ast=0.45m_0$ with $m_0$ as the free electron mass,
$k_0=10^8$ m$^{-1}$ and $B=1.5$ T.
It was reported that the value of the effective Lande g-factor for
heavy hole is highly anisotropic\cite{winkg}, we take its value
$g^\ast=7.2$ for GaAs system. 

The magnitude of Rashba strength ($\alpha$) 
depends explicitly on the external parameters\cite{rwink} like
electric field, detail of confinement and
hence can be tuned experimentally. 
So in this study, we take various values of $\alpha$ in such a way that the
corresponding length scale $l_\alpha=m^\ast \alpha/\hbar^2$ is of the order of 
few angstroms.

In Fig. (1) we show the time-evolution of the probability density
$\rho({\bf r},t)=\vert\Psi_1({\bf r},t)\vert^2$ + 
$\vert\Psi_2({\bf r},t)\vert^2$ of the heavy hole.
To do this we numerically evaluate the components of the wave packet 
$\Psi_1({\bf r},t)$ and $\Psi_2({\bf r},t)$ at a later time $t$
as described by Eqs. (\ref{wavefn1}) and (\ref{wavefn2}). 
The infinite series in Eqs. (\ref{wavefn1}) and (\ref{wavefn2}) converge
approximately when $n=30$ for $k_{0x}=1.5k_0$. However the convergence of 
these infinite series also depends on the value of $k_{0x}$. 
For larger $k_{0x}$ larger $n$ is required. We have checked that $n=60$ 
is appropriate for the convergence when $k_{0x}=2.5k_0$.
Figs. 1(a)-1(i) are plotted for 
$t=0, 2t_c, 4t_c, 5t_c, 6t_c, 7t_c, 8t_c, 9t_c$ and 
$10t_c$, respectively. Initially the wave packet is situated at 
the origin. We know that in a perpendicular magnetic field a charge particle
(without SOI) moves in a cyclotron orbit with time period 
$t_c=2\pi/\omega_c$. The radius of this circular orbit is 
$R_c= k_{0x} l_0^2$. 
The presence of SOI modifies $t_c $ and $R_c $ marginally.
In presence of the RSOI the hole wave packet
starts to diffuse asymmetrically around the cyclotron orbit while making circular 
motion 
as time goes on.
The diffused wave packet is making circular motion
with time period $\approx t_c$ and 
radius $\approx R_c$. 
At $t=4t_c$ [Fig 1(c)] the hole wave packet splits into two unequal 
peaks which are rotating with different velocities along the cyclotron orbit. 
The difference in the peak amplitudes nearly vanishes around 
$t=7t_c$ [Fig. 1(f)].
At the same time, the wave packet diffuses almost 
along the entire cyclotron orbit. Now the interference effect 
begins to occur between the tail and head of the diffused wave packet
and as a result a completely randomized pattern of hole wave packet
is observed after some more cycles as shown in Fig. 1(i). 
Recalling the case of two-dimensional electron system (2DES)
with linear RSOI in a 
perpendicular magnetic field where this kind of 
splitting of wave packet
occurs around $t \approx 45t_c$\cite{demi} for $B=1$ T. But in the present case 
of 2DHS this splitting occurs at much lesser time than the 2DES case. 
Moreover, the wave-packet does not diffuse
along the circular orbit in the case of 2DES with linear Rashba term.
These features can be attributed to the cubic Rashba term in the 2DHS.

\begin{figure}
\centering
\includegraphics[width=110mm]{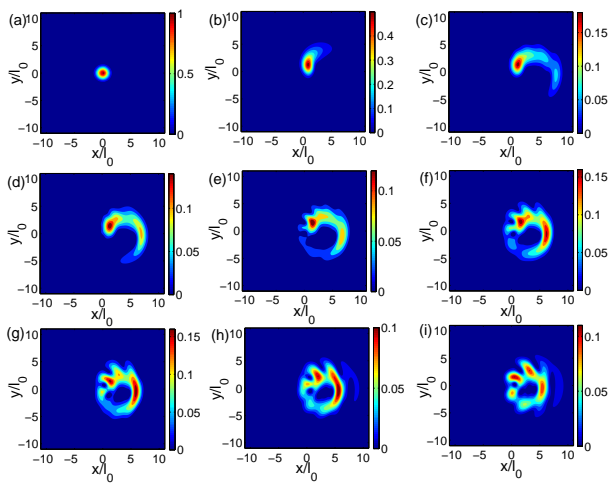}
\caption{(Color online) Time-evolution of the hole wave packet in 
a magnetic field at (a) $t=0$, (b) $t=2t_c$, (c) $t=4t_c$, 
(d) $t=5t_c$, (e) $t=6t_c$,
(f) $t=7t_c$, (g) $t=8t_c$, (h) $t=9t_c$, and (i) $t=10t_c$. 
In this case $\alpha$ is fixed in such a way that $l_\alpha=0.25$ nm.
In each plot the color bar represents the dimensionless 
parameter $\pi l_0^2\rho({\bf r},t)$.}
\end{figure}

\begin{figure}
\includegraphics[width=120mm]{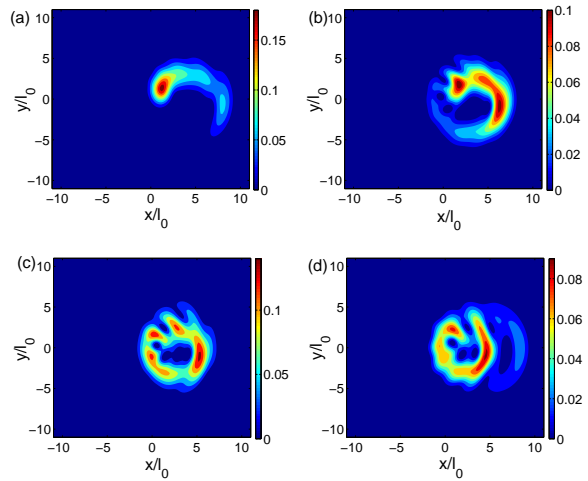}
\caption{(Color online) Time-evolution of the hole wave packet in 
a magnetic field at $t=4t_c$ for different values of 
$l_\alpha=m^\ast\alpha/\hbar^2$ namely (a) $l_\alpha=0.25$ nm, 
(b) $l_\alpha=0.35$ nm, (c) $l_\alpha=0.45$ nm, and 
(d) $l_\alpha=0.55$ nm. In this case magnetic field and 
the initial wave vector are fixed to $B=1.5$ T and 
$k_{0x}=1.5k_0$. In each plot the color bar represents the dimensionless 
parameter $\pi l_0^2\rho({\bf r},t)$.}
\end{figure}

In Fig. 2 spatial distribution of probability density at time 
$t=4t_c$ is shown. Different panels are plotted for different 
values of $\alpha$ such that $l_\alpha=0.25$ nm, $0.35$ nm, $0.45$ nm,
and $0.55$ nm. It is clear that as $\alpha$ increases the wave packet
diffuses very fast and covers the entire cyclotron orbit. 
The diffusion rate of the wave packet increases with increase of the
spin-orbit strength.
Also strong $\alpha$ helps to expedite the splitting and the randomization
of the hole wave packet.

\begin{figure}
\centering
\includegraphics[width=115mm]{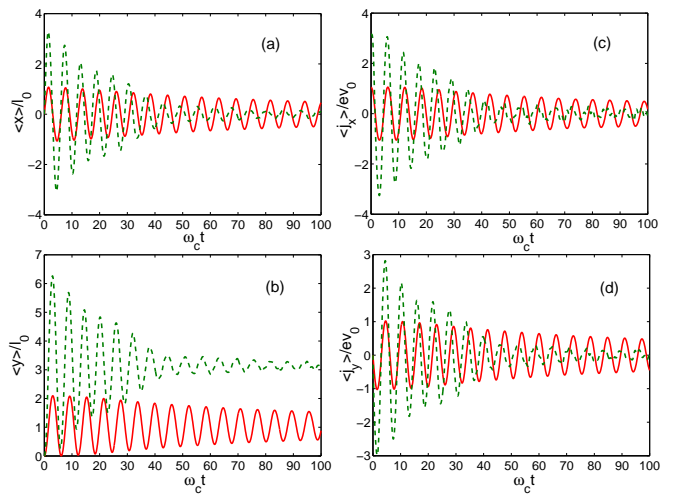}
\caption{(Color online) Time dependence of position and current components:
(a) $\la x\ra$, (b) $\la y\ra$, (c) $\la j_x\ra$, and (d) $\la j_y\ra$. 
Here in each panel solid and dashed lines correspond to 
$k_{0x}=0.5k_0$ and $1.5k_0$, respectively. Here $v_0$ is defined as 
$v_0=\hbar/(m^\ast l_0)$.}
\end{figure}

\begin{figure}
\centering
\includegraphics[width=110mm]{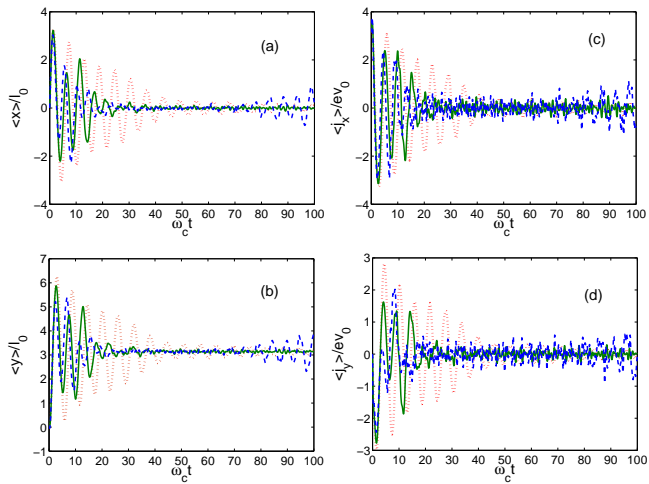}
\caption{(Color online) Time dependence of position and current components:
(a) $\la x\ra$, (b) $\la y\ra$, (c) $\la j_x\ra$, and (d) $\la j_y\ra$. 
Here in each panel dotted, solid and dashed lines correspond to 
$l_\alpha=0.25$, $0.5$, and $0.75$ nm, respectively. Here we define 
$v_0=\hbar/(m^\ast l_0)$.}
\end{figure}

In Fig. 3 we plot the average values of the position and current 
operators in $x$ and $y$ directions as a function of time $t$ for a 
fixed value of magnetic field $B=1.5$ T and fixed $\alpha$ such 
that $l_\alpha=0.25$ nm. Here, we define the current operators as 
$j_i(t)=ev_i(t)$ with $i=x$, $y$. In this case we consider two different values 
of initial wave vector $k_{0x}=0.5k_0$ (solid line)
and $k_{0x}=1.5k_0$ (dashed line). When $k_{0x}=0.5k_0$, 
oscillations appearing in $\la x\ra, \la y\ra, \la j_x\ra $ and 
$ \la j_y\ra$ are persistent in time. But in the case of higher 
$k_{0x}$, the amplitude of ZB 
decreases at later time which shows some kind of localization of
ZB oscillations.
This is because higher Landau levels are involving for 
higher values of $k_{0x}$. Comparing Figs. 3(a) and 3(b) 
we mention that $\la x\ra$ is oscillating about zero for both values of 
$k_{0x}$ whereas $\la y\ra$ is oscillatory but always positive because
Eq. (\ref{PosY}) contains a constant term. There is also
a definite phase difference between $\la x \ra$ and 
$\la y \ra$ and  $\la j_x \ra$ and $\la j_y \ra$ as clearly shown in Fig. 3.  
Investigating all the graphs in Fig. (3)
one can conclude that although initially there is no phase difference but 
an increment in $k_{0x}$ introduces a phase difference at later times.

Figure 4 describes the time dependence of position and current operators 
for a fixed magnetic field $B=1.5$ T and $k_{0x}=1.5k_0$. 
Different values of $\alpha$ such that $l_\alpha=0.25$ nm (dotted), 
$0.5$ nm (solid), and $0.75$ nm (dashed) have been considered.
It can be seen from Fig. 4 that the amplitude of ZB oscillation 
decreases as $\alpha$ increases. One interesting point is to be mentioned 
here that at stronger $\alpha$ ($l_\alpha=0.75$ nm) the ZB oscillations
start to reappear at large time.

\begin{figure}[t]
\centering
\includegraphics[width=110mm]{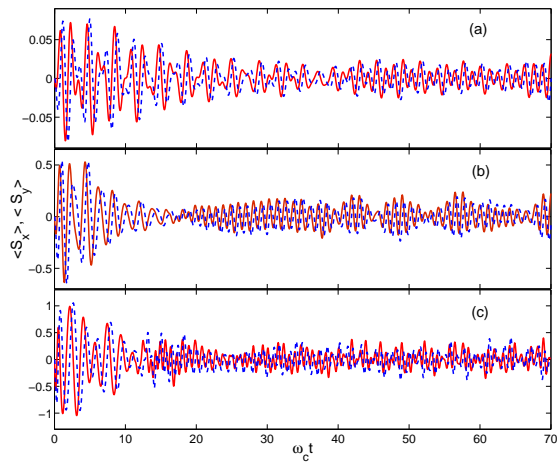}
\caption{(Color online)
Time dependence of the $x$ and $y$ components of the 
spin operator:
(a) $k_{0x}=0.5k_0$, (b) $k_{0x}=1.5k_0$, and (c) $k_{0x}=2.5k_0$.
Here in each panel solid and dashed lines represent $\la S_x\ra$ and $\la S_y\ra$,
respectively which are plotted in units of $\hbar$.}
\end{figure}

\begin{figure}[t]
\begin{center}\leavevmode
\includegraphics[width=110mm]{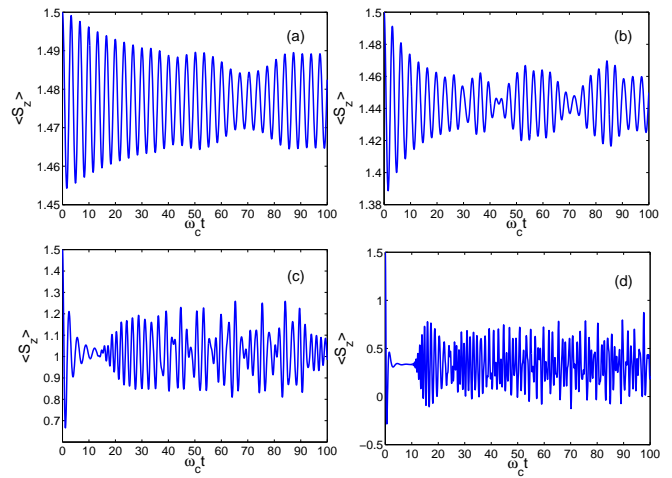}
\caption{(Color online)
Time dependence of the $z$ component of spin operator:
(a) $k_{0x}=0$, (b) $k_{0x}=0.5k_0$, (c) $k_{0x}=1.5k_0$, and 
(d) $k_{0x}=2.5k_0$.
Here in each panel $\la S_z\ra$ is plotted in units of $\hbar$.}
\label{Fig3}
\end{center}
\end{figure}

The time-dependent average values of the spin components are shown 
in Fig. 5 and 6. We consider various cases corresponding to the different 
values of the initial wave vector, namely $k_{0x}=0$, $0.5k_0$, $1.5k_0$ 
and $2.5k_0$ as mentioned in both figures. Figure 5 describes the variations 
of $\la S_x\ra$ and $\la S_y\ra$  with time $t$ for different values of $k_{0x}$. 
It can be seen that $\la S_x\ra$ and $\la S_y\ra$ maintains same oscillatory pattern
apart from a definite phase difference. An increase in $k_{0x}$ 
results a change in the oscillatory pattern significantly. The 
amplitude of the ZB oscillations, appearing in $\la S_x\ra$ and 
$\la S_y\ra$ increases as $k_{0x}$ increases. We have shown
the time dependence of $\la S_z\ra$ in Fig. 6. It should be noted 
that $\la S_z\ra$ does not vanish when there is no initial momentum
i.e. $k_{0x}=0$. This fact can be confirmed from Eq. (\ref{spinz}) 
which contains a non-zero term $H_{2n}(0)$. As $k_{0x}$ increases 
the oscillatory pattern of $\la S_z\ra$ changes abruptly. 
When $k_{0x}=0.5k_0$ a beating-like pattern appears in the 
oscillations of $\la S_z\ra$. From Fig. 6(b), (c), (d) it can be 
seen that the number of oscillations before the first beating node decreases
drastically as $k_{0x}$ increases. This complicated oscillatory 
pattern can be attributed to the fact that the infinite series 
in Eqs. (\ref{spinx}-\ref{spinz}) converges for higher 
values of $n$ as $k_{0x}$ increases. As a result more and more 
frequencies appear in the oscillations because of the involvement of 
higher Landau levels. Comparing Fig. 5 and 6 at $t=0$ 
$\la S_x\ra=0=\la S_y\ra$ where as 
$\la S_z\ra= 3\hbar/2$ and this is consistent with the fact that 
the initial wave packet is polarized along the $z$-direction.
The jittery motion in $S_z$ induces jittery motion
in $S_x$ and $S_y$.

\section{Summary}
In summary, we have studied wave packet propagation and $zitterbewegung$ 
of a heavy hole in III-V semiconductor quantum wells. 
We have visualized and discussed various consequences of the time-evolution
of the hole wave packet along the cyclotron orbit. 
The hole wave packet diffuses asymmetrically along the circular orbit while 
making cyclotron motion.
It is shown that the hole wave packet splits into two peaks with unequal 
amplitudes at certain time which depends on the spin-orbit interaction strength.
The two peaks rotate with different frequencies. 
The amplitude of the two peaks become nearly equal as time goes on.
After many cycles, tail and head of the diffused wave packet 
interfere with each other and produces a complete randomized pattern. 
The diffusion rate of the wave packet increases with increase of the
spin-orbit interaction strength.
Also strong spin-orbit coupling expedite the splitting and the randomization
of the wave packet.
Our results for the hole is compared with an electron in a linear Rashba system
in presence of the magnetic field. 
We have also studied ZB phenomenon in position, current and spin angular momentum
of the heavy hole.
The $zitterbewegung$ oscillations are very much sensitive to the
initial momentum of the wave packet and
the Rashba spin-orbit coupling constant.

\appendix
\section{Expectation values position and velocity operators}
In this appendix we shall present detailed derivation of 
the average values of various physical observables like 
position and velocity operators.

The calculation of the average value of the $y$ component 
of the position operator is easier than that of the $x$ component.
Let us first calculate average value of the $x$ component.
Equation (\ref{position}) can be written as 
$ \la x(t)\ra=\la x_1(t)\ra+\la x_2(t)\ra $,
where $\la x_i(t)\ra=\int dxdy\Psi_i^\ast({\bf r},t)x\Psi_i({\bf r},t)$.
The explicit form of $\la x_1(t)\ra$ is  
\begin{eqnarray}\label{posx1}
\la x_1(t) \ra & = & \frac{1}{2\pi^2l_0^2} 
\sum_{m,n=0}^\infty
\frac{\Gamma_{m+3}^\ast \Gamma_{n+3}}{2^{m+n}m!n!} 
\int d \Omega  e^{Q(x,y,u,u^\prime)} x  \nonumber\\
& \times & (-u)^n (-u^\prime)^m  H_n(y/l_0-u) 
H_m(y/l_0 - u^\prime),
\end{eqnarray}
where $ d \Omega = dxdydudu^\prime $, 
$Q(x,y,u,u^\prime)=\Lambda(x,y,u)+\Lambda^\ast(x,y,u^\prime)
=-ix(u^\prime-u)/l_0-F(y,u)-F(y,u^\prime) $ 
with $F(y,u)=(a-u)^2/2+u^2/4+(y/l_0-u)^2/2$.
The above equation [Eq. (\ref{posx1})]  can be re-written as 
\begin{eqnarray}\label{posx1n}
\la x_1 \ra & = & \frac{1}{2\pi^2l_0^2}\sum_{m,n}
\frac{\Gamma_{m+3}^\ast\Gamma_{n+3}}{2^{m+n}m!n!}\int dydudu^\prime 
e^{-F(y,u)-F(y,u^\prime)} \nonumber\\
& \times & (-u)^n(-u^\prime)^m H_n(y/l_0 - u) H_m(y/l_0 - u^\prime) 
\nonumber\\
& \times & \int dxxe^{-i x(u^\prime-u)/l_0}.
\end{eqnarray}

Using the standard results
\begin{eqnarray}\label{int1}
\int_{-\infty}^\infty dxxe^{-ix(u^\prime-u)/l_0}
=2\pi il_0^2 \frac{d}{du^{\prime}}[ \delta(u^\prime-u)]
\end{eqnarray}
and 
\begin{eqnarray}
\int du^\prime f(u^\prime) 
\frac{d}{du^{\prime}}\delta(u^\prime-u)
=-\frac{df(u^\prime)}{du^\prime} \Big|_{u^\prime=u},
\end{eqnarray}
Equation (\ref{posx1n}) becomes
\begin{eqnarray}\label{posxn3}
\la x_1\ra & = & - \frac{i}{\pi}\sum_{m,n}
\frac{\Gamma_{m+3}^\ast\Gamma_{n+3}}{2^{m+n}m!n!}\int dydu (-u)^{m+n}
e^{-2F(y,u)}\nonumber\\
& \times & H_n(y/l_0 - u)\mu_m(u).
\end{eqnarray}
Here, $\mu_m(u)=(m/u - 5u/2 + a + y/l_0) H_m (y/l_0 - u)
-2mH_{m-1}(y/l_0 - u)$.

Integrating over $y$ variable in Eq. (\ref{posxn3}) using the
properties of the Hermite polynomials, we finally get
\begin{eqnarray}\label{posxn4}
\la x_1\ra & = & -\frac{i}{\pi}\exp(-a^2/3) 
\sum_{m,n}
\frac{(-1)^{m+n}\Gamma_{m+3}^\ast\Gamma_{n+3}}{2^{m+n}m!n!}\nonumber\\
&\times& \int du I_{mn}(u) u^{m+n} e^{-(\sqrt{3/2}u-\sqrt{2/3} a)^2},
\end{eqnarray}
where $I_{mn}(u)$ is given by 
\begin{eqnarray}
I_{mn}(u) & = & l_0 \sqrt{\pi}2^nn!\Big\{\Big(\frac{m}{u}-\frac{3u}{2}+a\Big)
\delta_{m,n} + \frac{1}{2}\delta_{m, n-1}\nonumber\\
&+&(n+1-2m)\delta_{m,n+1}\Big\}.
\end{eqnarray}

Again, integrating over $u$ variable in Eq. (\ref{posxn4}) 
using the following result
\begin{eqnarray}
\int_{-\infty}^\infty dw w^ne^{-(w-b)^2}=\sqrt{\pi}\frac{H_{n}(ib)}{(2i)^n},
\end{eqnarray}
finally we have, 
\begin{eqnarray}
\la x_1(t)\ra&=&\frac{l_0}{3}\exp(-a^2/3)\sum_{n=0}^\infty
\frac{i(-1)^{n+1}}{n!(12)^n}\nonumber\\
&\times&\Im(\Gamma_{n+4}^\ast\Gamma_{n+3})
H_{2n+1}(i\sqrt{2/3}a).
\end{eqnarray}

In the similar way, $\la x_2(t)\ra$ is obtained as 
\begin{eqnarray}
\la x_2(t)\ra&=&\frac{l_0}{3}\exp(-a^2/3)\sum_{n=0}^\infty
\frac{i(-1)^{n+1}}{n!(12)^n}\sqrt{\frac{n+4}{n+1}}\nonumber\\
&\times&\Im(\Delta_{n+4}^\ast\Delta_{n+3})H_{2n+1}(i\sqrt{2/3}a).
\end{eqnarray}  

Similar to the calculation of $\la x(t) \ra$, one can calculate
$\la y(t) \ra$.

To calculate $ \la {\bf v}(t) \ra $, we need to know the matrix elements
$ \la {\bs \Pi}(t) \ra, \la \Psi_1 |\Pi_{-}^2| \Psi_2 \ra $ and its complex
conjugate. Following the above mentioned method, we have
\begin{eqnarray}
\left(
\begin{array}{c}
\la \Pi_x(t) \ra \\
\la \Pi_y(t) \ra \\
\end{array}\right)
& = & \frac{\hbar}{3l_0}\exp(- a^2/3)
\left(
\begin{array}{c}
\mbox{Re} Z(t) \\
- \mbox{Im} Z(t) \\
\end{array}\right),
\end{eqnarray}
where $Z(t)$ is given by
\begin{eqnarray}
Z(t) &=&
\sum_{n=0}^\infty \frac{i(-1)^{n+1}}{n!(12)^n} 
H_{2n+1}(i\sqrt{2/3}a) \nonumber\\
& \times &
\Big\{\Gamma_{n+3}^\ast\Gamma_{n+4}
+ \sqrt{\frac{n+4}{n+1}}
\Delta_{n+3}^\ast\Delta_{n+4}\Big\},
\end{eqnarray}

and

\begin{eqnarray}\label{velint1}
&&\int\Psi_1^\ast({\bf r},t)\Pi_{-}^2\Psi_2({\bf r},t)dxdy=
\frac{\sqrt{2}\hbar^2}{3l_0^2}\exp(-a^2/3)
\times \nonumber\\
&{}& \sum_{n=0}^\infty
\frac{i(-1)^{n}}{n!(12)^n} \sqrt{\frac{(n+2)(n+3)}{n+1}}
\Gamma_{n+4}^\ast\Delta_{n+3}
H_{2n+1}(i\sqrt{2/3}a). \nonumber\\
\end{eqnarray}

\end{document}